\documentclass[prd,nofootinbib,floatfix,aps,preprint]{revtex4-1}
\usepackage{amsmath}
\usepackage{amsfonts}
\usepackage[linktoc=all]{hyperref}
\usepackage{cleveref}
\usepackage{color}

\newcommand{\Mp}{M_\mathrm{Pl}}
\newcommand{\Meff}{M_\mathrm{eff}}
\newcommand{\dd}{\mathrm{d}}
\newcommand{\sdg}{\sqrt{-g}}
\newcommand{\sdgt}{\sqrt{-\tilde g}}
\newcommand{\mn}{{\mu\nu}}
\newcommand{\ab}{{\alpha\beta}}
\newcommand{\Lg}{\lambda_\mathrm{g}}
\newcommand{\mo}{\mathcal{O}}

\allowdisplaybreaks

\makeatletter
\DeclareRobustCommand{\rcite}[1]{%
  \rcite@aux#1,\@nil{#1}%
}
\def\rcite@aux#1,#2\@nil#3{%
  \if\relax#2\relax
    Ref.~\cite{#3}%
  \else
    Refs.~\cite{#3}%
  \fi
}
\makeatother

\begin{document}

\count\footins = 1000 

\title{Higher-derivative operators and effective field theory for general scalar-tensor theories}
\author{Adam R. Solomon}
\author{Mark Trodden}
\affiliation{Center for Particle Cosmology, Department of Physics and Astronomy, University of Pennsylvania, 209 S. 33rd St., Philadelphia, PA 19104, USA}
\date{\today}

\begin{abstract}
We discuss the extent to which it is necessary to include higher-derivative operators in the effective field theory of general scalar-tensor theories. We explore the circumstances under which it is correct to restrict to second-order operators only, and demonstrate this using several different techniques, such as reduction of order and explicit field redefinitions. These methods are applied, in particular, to the much-studied Horndeski theories. The goal is to clarify the application of effective field theory techniques in the context of popular cosmological models, and to explicitly demonstrate how and when higher-derivative operators can be cast into lower-derivative forms suitable for numerical solution techniques.
\end{abstract}

\maketitle

\tableofcontents

\section{Introduction}

Modern physical theories are typically understood through the lens of effective field theory (EFT). The EFT philosophy of model-building is straightforward: physics at a given energy scale generally does not require a detailed understanding of physics at much higher energies, allowing us to model the lower-energy phenomena with an \emph{effective} theory suited to those scales. This principle, known as decoupling, underlies our ability to describe fluid dynamics without knowing the Standard Model, for instance, or to test general relativity in the absence of a theory of quantum gravity. To construct models in the EFT approach, we parametrize our ignorance about more fundamental physics by writing down the most general Lagrangian consistent with a given particle content and set of symmetries. This will typically take the form of a finite number of relevant and marginal operators, as well as a tower of irrelevant operators suppressed by some energy scale $\Lambda$, taken to be the scale around which the EFT breaks down and at which new ultraviolet (UV) physics becomes crucial.\footnote{Strictly speaking $\Lambda$ is where \emph{perturbative} unitary breaks down; usually this is where new physics enters (i.e., non-perturbative unitarity breaks down), though there are cases where new physics enters in at a scale high above $\Lambda$ \cite{deRham:2014wfa}.} When experiments are performed at energies $E$ well below $\Lambda$, only a finite number of these irrelevant operators are probed by the results, with the precision of the experiment telling us to which order in $E/\Lambda$ we need to work.

EFTs are the setting for practically all of physics at the energy scales presently accessible to experiment. In particular, both the Standard Model and general relativity are effective field theories, expected to be the low-energy descriptions of some UV completion.\footnote{For the Standard Model as an EFT, see, e.g., \rcite{Grzadkowski:2010es,Buchmuller:1985jz}. For reviews of the EFT description of gravity, see \rcite{Donoghue:1994dn,Burgess:2003jk}.} In particular, the famous non-renormalizability of general relativity is not a problem for low energy, low curvature phenomena, when the theory is viewed as an EFT; it is simply the lowest-dimension operator in some (potentially infinite) expansion that is only valid up to some energy scale, and there is no reason to expect it to be renormalizable to arbitrarily high energies. Most of the modifications of general relativity in the literature are similarly non-renormalizable,\footnote{Including curvature-squared terms to general relativity improves its renormalizability \cite{Stelle:1976gc}, but this theory contains a ghost.} and so should also be seen only as effective theories.

Another somewhat orthogonal approach to theory building, common in cosmology for example, is to enumerate all possible actions, for some given field content, whose equations of motion are second order. This is motivated by a desire to avoid the Ostrogradski instability: when the equations of motion are higher than second-order in time derivatives, the Hamiltonian is almost always unbounded from below, leading to catastrophic instability.\footnote{For the small number of exceptions to this rule, see \rcite{Woodard:2006nt}.} Demanding the absence of this instability has led to the discovery of a zoo of interesting theories which are second-order in both the action and in the equations of motion, with an underlying structure causing the third- and fourth-derivative terms arising from variation to cancel out.

Consider, for example, a scalar field on flat space, and include in its action terms with (at least) two derivatives per field. Most of these terms lead to fourth-order equations of motion: for example, the term $(\Box\phi)^2$ leads, upon variation, to $\Box^2\phi$. The equation of motion is then fourth-order in time, and requires four initial conditions rather than the usual two. Ostrogradski's theorem tells us that this extra degree of freedom inevitably leads to an exponential instability. However, some terms buck this logic: for example, the \emph{cubic galileon}, $(\partial\phi)^2\Box\phi$, contributes to the equation of motion the strictly second-order term $(\partial_\mu\partial_\nu\phi)^2-(\Box\phi)^2$.

The cubic galileon is one of a class of scalar-field Lagrangians (sometimes referred to as \emph{generalized galileons}) which lead to second-order equations of motion \cite{Nicolis:2008in,Deffayet:2011gz,Padilla:2010de,Hinterbichler:2010xn,Goon:2011uw,Goon:2011qf,Goon:2011xf}. Similar stories have been spelled out for pure gravity, with the most general set being the so-called \emph{Lovelock} terms \cite{Lovelock:1971yv},\footnote{In four dimensions, the only such term is the Einstein-Hilbert term.} while if the metric is coupled to a scalar, one similarly finds the \emph{Horndeski} terms \cite{Horndeski:1974wa,Deffayet:2011gz}, which contain the generalized galileons in the flat-space limit and have in recent years found widespread use in cosmological applications, e.g., to explain dark energy or inflation.\footnote{See \rcite{Joyce:2014kja} and references therein.}

When considering new theories of, for instance, gravity, the claim is often made that the most general healthy theory is therefore of the Lovelock or Horndeski type, or one of their extensions to include different fields. However, from the EFT point of view the absence of higher derivative operators is not a requirement, and indeed if it were necessary---if, for example, scalar-tensor EFTs were required to be in the Horndeski class---this would be a surprising and interesting restriction on the space of viable effective theories \cite{Burgess:2014lwa}. These statements are reconciled by the fact that the Ostrogradski ghost that appears in EFTs with higher-derivative equations of motion is unphysical: its mass is typically around the cutoff, meaning that processes that are well-described by the EFT cannot possibly excite it and induce the dreaded runaway instability. Nevertheless the equations of motion that appear in such theories are clearly not fit to be simply solved as-is, because those solutions would exhibit this (unphysical) instability. This is a problem in particular when numerically solving the equations of motion in EFTs with higher-derivative operators.

It is our goal to discuss the proper way (or ways) to treat these EFTs and to point a way forward for the use of proper EFTs in confrontation with experiment and observation, with a particular focus on modified gravity. We aim to understand the circumstances under which we need to take into account higher-derivative terms, the forms that such terms take, and the criteria for them to become important in comparison to strictly-second-order terms. We hope that this will help to clarify some of the confusions that sometimes arise when discussing these types of theories and also be of practical help to those working on numerically solving the relevant equations of motion.

What this amounts to is a formulation of how to treat modified gravity as an EFT in much the same way as physics beyond the Standard Model is already understood: identifying the scale $\Lambda$ at which the first new physics enters, and using experiment to measure or bound the coefficients of the various EFT operators. The question of higher derivatives is less pressing in the Standard Model EFT; at dimension-6 there are already 84 operators, none of which are higher-derivative in nature \cite{Henning:2015alf}, leaving plenty of targets for experiment without worrying about the question of exorcising ghosts. However, as we will see, for theories of modified gravity, which have a much simpler particle content and therefore far fewer EFT operators, the problem of higher derivatives confronts us quite early.

We expect that different audiences will find different parts of this paper useful, and so for reference we will briefly summarize each section.

In \cref{sec:higher-derivs} we review higher derivatives in effective field theories, demonstrating why they are not problematic when considered from the EFT point of view and showing an explicit example of healthy UV physics which lead to an apparently unhealthy EFT. We make the well-known argument (following, e.g., \rcite{Simon:1990ic,Burgess:2014lwa}) that not all of the solutions to such an EFT are physical, in the sense that the problematic runaway solutions cannot be expected to accurately capture the full solutions to the UV theory. We then provide in \cref{sec:ghostbusting} an overview of methods to extract these physical solutions given a higher-derivative EFT. To the best of our knowledge, most of this section is review and as such is treated pedagogically.

In \cref{sec:scalar-eft} we take a deeper look at how one finds these physical solutions in practice for EFTs of a real scalar field. We consider a top-down example and find that the quartic galileon describes physical solutions to the low-energy dynamics of the Goldstone mode for a complex scalar with spontaneously-broken U(1) symmetry. We then construct a bottom-up operator basis for a single scalar field with shift symmetry, with a focus on eliminating higher-derivative operators through integrations by parts and field redefinitions. We find that all such operators up to mass dimension 11 can be removed. This somewhat-surprising result demonstrates that, even if one should not restrict to operators with second-order equations of motion \textit{a priori}, higher-derivative terms might not affect the physical solutions until a rather high order in the EFT expansion. This provides at least a partial justification for the use of these second-order terms, in particular (in the scalar field context) the generalized galileons.

Finally, in \cref{sec:mg-eft} we apply these techniques to the more physically-relevant case of scalar-tensor theories of modified gravity, from both the bottom-up and top-down approaches. From the bottom up, we construct an operator basis up to six-derivative order, finding seven non-Horndeski terms. We argue that these terms should be considered alongside their Horndeski counterparts in cosmological applications.\footnote{EFT approaches to cosmology taking into account higher-derivative interactions using the methods discussed in this paper have previously been considered for quantum corrections to gravity \cite{Parker:1993dk}, during inflation \cite{Weinberg:2008hq}, and for dark energy \cite{Bloomfield:2011np}.} We also consider an example top-down scalar-tensor EFT with a known UV provenance, showing in that case that, analogously to \cref{sec:u1-redefs}, the physical solutions can be obtained solely with Horndeski terms. We conclude in \cref{sec:conc}.

We would like the reader to leave understanding that higher derivatives are generic in EFTs, as is well-known in the context of particle physics theories. In particular, healthy UV physics does \emph{not} imply a ghost-free EFT. The restriction to theories with strictly second-order equations of motion, such as Lovelock or Horndeski, is too strong a restriction in the absence of any further justification. Instead non-renormalizable theories, such as are used in, e.g., modified gravity, require treating as effective field theories, performed in a derivative expansion and with higher derivative terms allowed, to be treated as we discuss below. For the specific case of scalar-tensor gravity we explicitly perform the construction of such an EFT, identifying the most relevant non-Horndeski terms which should be considered when comparing to observations.

\section{Higher derivatives in EFTs}
\label{sec:higher-derivs}

\subsection{A quantum gravity warm-up}
\label{sec:qg}

We begin with a brief historical diversion, which we hope will make clear that, in the EFT context, higher derivatives should not be taken entirely seriously. It begins by asking a simple question: is flat space stable in quantum gravity? One would hope that the answer is a resounding ``yes": after all, we believe that there should exist a theory of quantum gravity with general relativity as its low-energy limit, and similarly undergirding much of our day-to-day lives is the belief that flat space is not unstable to catastrophic and nearly instantaneous decay.

General relativity can be well thought of as the lowest-energy piece of an EFT \cite{Donoghue:1994dn,Burgess:2003jk}; indeed, this fact underlies our ability to use general relativity (both classical and semiclassical) in the first place, despite it not being renormalizable. We simply see it as being valid up to some cutoff $\Lambda$, above which new physics, perhaps quantum gravity, becomes dominant. At energies below $\Lambda$, we can express the action as a series in powers of $1/\Lambda$ involving all possible (diffeomorphism-invariant) terms we can write down involving the metric,
\begin{equation}
S = \int\dd^4x\left[\frac{\Mp^2}{2}R + \alpha R^2 + \beta R_\mn R^\mn + \frac{1}{\Lambda^2} \mathrm{Riem}^3 + \cdots\right] \ ,
\end{equation}
where in the last term $\mathrm{Riem}^3$ refers to all the various contractions of three curvature tensors, $\alpha$ and $\beta$ are dimensionless constants, and we have left out a possible $R_{\mn\ab}R^{\mn\ab}$ term because the Gauss-Bonnet combination $G=R^2-4R_\mn R^\mn+R_{\mn\ab}R^{\mn\ab}$ is a total derivative. As we know from Lovelock's theorem \cite{Lovelock:1971yv,Lovelock:1972vz}, precisely every term in this series besides the Einstein-Hilbert term will lead to higher-order equations of motion. In particular, terms like $R_\mn R^\mn$ violate Ostrogradski's theorem and thus lead to runaway instability.\footnote{Terms like $R^2$ do not violate Ostrogradski due to a loophole, although they do unlock an extra scalar degree of freedom contained in the metric \cite{Woodard:2006nt}.} Does this mean, as had been suspected once upon a time \cite{Horowitz:1978fq,Horowitz:1980fj,Hartle:1981zt}, that quantum gravity corrections inevitably destabilize flat space? The answer, as shown explicitly by Simon \cite{Simon:1990jn}, fortunately turns out to be no.\footnote{Indeed, the quadratic terms do not have \emph{any} effect on metric perturbations in vacuum \cite{Simon:1990jn,Simon:1991bm,Flanagan:1996gw}. As we will see below, this is because the $R^2$ and $R_\mn^2$ vanish on-shell and can therefore be removed entirely with a perturbative field redefinition.} In the rest of this section it will become clear why such terms do not lead to these instabilities.

\subsection{Healthy UV theories can have na\"ively unhealthy EFTs}
\label{sec:u1}

In this subsection we will review an example used in \cite{Burgess:2014lwa} to explicitly show that healthy UV physics can lead to a low-energy EFT which appears to contain a ghost. This point has certainly been known for a long time, but is perhaps not universally appreciated.

Consider a complex scalar $\Phi$ with a spontaneously-broken U(1) symmetry. The action is
\begin{equation}
S = \int\dd^4x\left[-\partial_\mu\Phi^\star\partial^\mu\Phi - V(\Phi^\star\Phi)\right] \ ,
\end{equation}
with the potential given by
\begin{equation}
V = \frac\lambda2\left(\Phi^\star\Phi-\frac{v^2}{2}\right)^2 \ .
\end{equation}
The action is symmetric under $\Phi\to e^{i\omega}\Phi$, but the symmetry is spontaneously broken in the vacuum $\Phi^\star\Phi=v^2/2$. The spectrum of fluctuations about this solution contains a massive mode with $M^2=\lambda v^2$ and a massless Goldstone boson. This can be made explicit by splitting up $\Phi$ into two real fields $\rho$ and $\theta$,
\begin{equation}
\Phi = \frac{v}{\sqrt{2}}\left(1+\rho\right)e^{i\theta} \ .
\end{equation}
In terms of these variables the action is that of two coupled scalar fields,
\begin{equation}
\frac{S}{v^2} = \int\dd^4x\left[-\frac12(\partial\rho)^2 - \frac12(1+\rho)^2(\partial\theta)^2 - V(\rho)\right] \ ,
\end{equation}
with
\begin{equation}
V(\rho) = \frac{M^2}{2}\left(\rho^2+\rho^3+\frac14\rho^4\right) \ .
\end{equation}
The equations of motion for $\rho$ and $\theta$ are
\begin{align}
\Box\rho-(1+\rho)(\partial\theta)^2-V'&=0 \ ,\\
\partial_\mu\left[(1+\rho)^2\partial^\mu\theta\right]&=0 \ .
\end{align}

As discussed in more detail in \cite{Burgess:2014lwa}, we can integrate out the heavy field $\rho$ by solving its equation of motion and inserting this solution back into the action to obtain a theory for $\theta$ alone. In general the solution for $\rho$ is highly non-local; following \cite{Burgess:2014lwa} we can write it recursively as
\begin{equation}
\rho(x) = -\int\dd^4x'G(x,x')\left\{\left[1+\rho(x')\right]\left[\partial\theta(x')\right]^2+V_\mathrm{int}'\left[\rho(x')\right]\right\} \ , \label{eq:rhosol}
\end{equation}
where $V_\mathrm{int} = V-\frac12M^2\rho^2$ and the Green's function $G(x,x')$ is given by $(-\Box+M^2)G(x,x')=\delta^{4}(x-x')$.

This expression for $\rho$ is an utter mess, and the effective action we obtain for $\theta$ through it is useless. The key point of the effective field theory logic is that for energies well below the mass of $\rho$, $E\ll M$, the $\Box\rho$ part of the equation of motion becomes subdominant and so we can write $\rho$ as an expansion in powers of $1/M$,
\begin{equation}
\rho = \displaystyle\sum_{n=1}^{\infty}\frac{r_n}{M^{2n}} \ ,
\end{equation}
and by truncating at some finite order in $1/M$ the solution \eqref{eq:rhosol} for $\rho$, and therefore the effective action for $\theta$, \emph{becomes} local. This is precisely the point of using an effective field theory: given a small parameter, like $E/M$, we expand in that parameter as early as possible to maximally simplify subsequent calculations.

It is precisely this step where a healthy (albeit non-local) theory for $\theta$ turns into an apparently-unhealthy one. Working to $\mo(M^{-4})$ one finds
\begin{equation}
\rho = -\frac{(\partial\theta)^2}{M^2} - \frac{(\partial\theta)^4+2\Box(\partial\theta)^2}{2M^4} + \mathcal{O}\left(\frac{1}{M^6}\right)
\end{equation}
and the effective action for $\theta$ reads
\begin{equation}
\frac{S}{v^2} = \int\dd^4x\left[-\frac12(\partial\theta)^2 + \frac{1}{2M^2}(\partial\theta)^4 - \frac{2}{M^4}\theta_\mn\theta^{\mu\rho}\theta^\nu\theta_\rho + \mo\left(\frac{1}{M^6}\right)\right] \ , \label{eq:cliff-eft-action}
\end{equation}
where we have defined $\theta_\mu\equiv\partial_\mu\theta$, and $\theta_\mn\equiv\partial_\mu\partial_\nu\theta$.

At $\mo(M^{-2})$ we have a perfectly healthy term of the so-called $P(X)$ class. However, at the next order, disaster seems to strike: this term contributes third- and fourth-order derivatives of $\theta$ to the equation of motion, and therefore leads to runaway instability. The UV theory clearly is healthy, but the low-energy effective action is plagued by an Ostrogradski ghost!

What happened here? The point is that by truncating at finite order in $1/M$, our nonlocal mess becomes local, but at the price of introducing higher derivatives. If we were to resum all the operators at $\mo(M^{-8})$ and higher, we would find the original, ghost-free theory. The EFT does have a ghost, but the mass of the ghost is $\sim M$, so we cannot produce it while remaining within the r\'egime of validity of the EFT: if we were to start probing energies high enough that we would worry about producing this ghost, we would not be able to limit ourselves to the three terms in \cref{eq:cliff-eft-action}, but would need to know \emph{all} of the infinite terms in the expansion. These terms cure the instability, as they simply resum to the ghost-free theory we started with.

The above story is morally analogous to more complicated real-world cases such as physics beyond the Standard Model or general relativity. We aim to use the EFT framework to allow our measurements to properly guide the search for a more fundamental theory. By restricting to theories like Horndeski, as is frequently done in, e.g., the cosmological context, one runs the very real risk of similarly missing physics which has every right to be there.

\subsection{Methods for exorcising ghosts}
\label{sec:ghostbusting}

Having established that ghosts in EFTs are perfectly consistent with the existence of a healthy, ghost-free UV origin, we now review a number of different techniques to treat these higher-derivative equations of motion, discuss the utility of each in real-world calculations, and demonstrate that they are complementary. There will be a general progression from simplest to most useful.

The important point is that when an EFT possesses a higher-derivative equation of motion, only a subset of its solutions correctly reflects the solutions to the full UV theory.\footnote{Equivalently, such a theory is not just defined by its Lagrangian, but implicitly contains additional, perturbative constraints which arise by demanding convergence \cite{Jaen:1986iz,Eliezer:1989cr,Simon:1990jn}.} As an extremely simple example, consider a one-dimensional point particle with the Lagrangian \cite{Woodard:2006nt,Burgess:2014lwa}
\begin{equation}
L = \frac12\dot{x}^2 - \frac12\epsilon \ddot{x}^2 - \frac12m^2x^2 + \mo(\epsilon^2) \ , \label{eq:ppaction-m}
\end{equation}
where $\epsilon$ is a positive expansion parameter\footnote{The restriction to positive $\epsilon$ is for simplicity. If we have $\epsilon<0$ then the solution \eqref{eq:pp-sol} is oscillatory rather than exponentially growing and decaying. However, the Ostrogradski instability is nevertheless present as the Hamiltonian is unbounded from below; see \rcite{Woodard:2006nt} for a more detailed discussion.} analogous to $M^{-2}$ in the previous subsection, and is taken to be small in the sense that $\epsilon \dd^2/\dd t^2\ll 1$. The equation of motion is fourth-order,
\begin{equation}
\ddot x + m^2 x - \epsilon x^{(4)}=\mo(\epsilon^2) \ . \label{eq:ppeom}
\end{equation}
We have chosen this very simple example because it can be solved exactly,
\begin{equation}
x = A_+e^{k_+t} + B_+e^{-k_+t} + A_-e^{k_-t} + B_-e^{-k_-t} \ , \label{eq:pp-sol}
\end{equation}
where $A_\pm$ and $B_\pm$ are constants to be fixed by (four) initial conditions, and
\begin{equation}
k_\pm^2 = \frac{1}{2\epsilon}\left(1 \mp \sqrt{1+4\epsilon m^2}\right) \ .
\end{equation}
The terms with $k_+$ in the exponent are oscillatory, but the terms with $k_-$ in the exponent are not, as can be seen by expanding $k_\pm$ to leading order in $\epsilon$,
\begin{equation}
k_+\approx im + \mo(\epsilon),\qquad k_- \approx \frac{1}{\sqrt\epsilon} + \mo(\epsilon^0) \ .
\end{equation}
We see clearly the effects of the ghost. In contrast to the $\epsilon=0$ case, which is solved by $x = A_+e^{imt} + B_+e^{-imt}$, we now have two more initial conditions, $A_-$ and $B_-$, and moreover these are associated with runaway exponentially growing and decaying solutions. This is of course bad behavior, but it is also not physical: the exponentials $e^{\pm t/\sqrt\epsilon}$ are not consistent with a perturbative expansion in $\epsilon$. Only the solutions with $A_-=B_-=0$ are \emph{physical} insofar as they have any hope of perturbatively reflecting solutions of the full UV theory.

This example is simple, but serves to illustrate precisely what is going on when higher derivatives appear in EFTs. The space of solutions to the higher-order equation(s) of motion is larger than the space of solutions we expect, because the higher-derivative terms (which are supposed to be perturbative corrections to the lowest-order equation) raise the derivative order and therefore require us to specify extra initial conditions. These new initial conditions are attached to (usually unhealthy) solutions which do not properly reflect the UV physics. The solution space contains a subspace of \emph{physical} solutions, and it is these solutions that need to be identified when confronting such an EFT against data. In the following subsections we discuss various equivalent methods of identifying such solutions, with varying degrees of utility.

\subsubsection{Explicitly solving the equations of motion}

In the simple point particle example above, we were able to solve the equation of motion explicitly and identify solutions which were not perturbative in $\epsilon$. By doing so we identified the correct choice of initial conditions to eliminate the unphysical solutions. There may be other situations in which such a procedure can be done.

\subsubsection{Solving perturbatively}

An EFT is supposed to reflect the UV theory order by order in the expansion parameter $\epsilon$, so solutions that correctly do this can be identified by writing the solutions in such an expansion, e.g.,
\begin{equation}
x = x_0 + \epsilon x_1 + \mo(\epsilon^2)
\end{equation}
in the point particle example, or
\begin{equation}
\theta = \theta_0 + \frac{1}{M^2}\theta_1 + \frac{1}{M^4}\theta_2 + \mo(M^{-6})
\end{equation}
in the complex scalar example.\footnote{For some limitations of this approach, see the discussion in \rcite{Flanagan:1996gw}.} For example, solving the point-particle equation of motion \eqref{eq:ppeom} perturbatively we find
\begin{align}
&&&&&&\mo(\epsilon^0): && \ddot x_0 + m^2x_0^2 &= 0 &&&&&&\\
&&&&&&\mo(\epsilon^1): && \ddot x_1 + m^2x_0^2 - x_0^{(4)} &= 0 \ . &&&&&&
\end{align}
Solving these we obtain, to $\mo(\epsilon)$,
\begin{equation}
x = A\cos(mt)+ B\sin(mt)+ \frac12\epsilon m^3t\left[A\sin(mt)-B\cos(mt)\right]+\mo(\epsilon^2) \ . \label{eq:pertsol}
\end{equation}
Note that when solving for $x_1$, two extra constants of integration appear, but these are spurious and can be removed by perturbatively shifting $A$ and $B$. The perturbative solution \eqref{eq:pertsol} is precisely what we would obtain from the exact solutions \eqref{eq:pp-sol} by removing the runaway modes, i.e., setting $A_-=B_-=0$, and expanding to $\mo(\epsilon)$. This agreement between the full solution with runaways removed and the perturbative solution is not specific to this example, and holds for a wide range of particle and field theories \cite{Eliezer:1989cr}.

\subsubsection{Reduction of order}
\label{sec:reduction}

Given an EFT with higher-order equations of motion, the fact that the offending terms appear at subleading orders in an expansion in a small parameter can be used to \emph{reduce the order} of the equation \cite{Simon:1990jn,Parker:1993dk}. More precisely, we can use the perturbative nature of the EFT to obtain a second-order equation of motion which is equivalent up to a given order in the EFT expansion. Continuing with the point particle example for illustration, consider again the equation of motion \eqref{eq:ppeom},
\begin{equation}
\ddot x + m^2 x - \epsilon x^{(4)}=\mo(\epsilon^2) \ .
\end{equation}
Notice that we have explicitly put $\mo(\epsilon^2)$ on the right-hand side, rather than 0, to make explicit that the equation of motion can only be trusted up to that order in $\epsilon$. This means that we can multiply by $\epsilon$ and rearrange to find
\begin{equation}
\epsilon\ddot x + m^2 \epsilon x=\mo(\epsilon^2) \ .
\end{equation}
Taking two derivatives of this we obtain
\begin{equation}
\epsilon x^{(4)} + m^2 \epsilon \ddot x=\mo(\epsilon^2) \ .
\end{equation}
We can add this to the original equation of motion to find 
\begin{equation}
\left(1+\epsilon m^2\right)\ddot x + m^2 x=\mo(\epsilon^2) \ , \label{eq:ppeom-red}
\end{equation}
or, equivalently, after multiplying through by $(1-\epsilon m^2)$,
\begin{equation}
\ddot x + \tilde m^2 x=\mo(\epsilon^2) \ , \label{eq:redeom}
\end{equation}
where
\begin{equation}
\tilde m^2 \equiv \left(1-\epsilon m^2\right)m^2 \ . \label{eq:mtilde}
\end{equation}

This procedure reduced the order of the equation of motion from fourth to second. Of course, one cannot ordinarily reduce the order of a differential equation just with these kinds of simple manipulations: the original higher-derivative term is still present, but it is now at $\mo(\epsilon^2)$ rather than $\mo(\epsilon)$, and we have set up our EFT so that we are ignoring terms at $\mo(\epsilon^2)$ and higher. We have shuffled the problematic term off to a higher order where we can happily ignore it. From a physical standpoint this is perfectly sensible: we are assuming (as explicitly demonstrated in the complex scalar example in \cref{sec:u1}) that the higher-order terms we are ignoring cure the instability anyway.

The main utility of this procedure is that we can solve this reduced-order equation \emph{exactly} and still trust that we are probing the physical solutions of the EFT. For example, solving \cref{eq:redeom} and treating the $\mo(\epsilon^2)$ on the right-hand side as exactly zero, we have
\begin{equation}
x = A\cos(\tilde m t) + B\sin(\tilde m t) \ .
\end{equation}
Using \cref{eq:mtilde} to express this in terms of the original $m$ and expanding to $\mo(\epsilon)$, we find
\begin{equation}
x = A\cos(mt)+ B\sin(mt)+ \frac12\epsilon m^3t\left[A\sin(mt)-B\cos(mt)\right]+\mo(\epsilon^2) \ .
\end{equation}
This is precisely the solution \eqref{eq:pertsol} we obtained above by solving the higher-order equation of motion perturbatively, and which we could also have obtained by solving the higher-order equation exactly, throwing away the runaways, and expanding to $\mo(\epsilon)$.

In the field-theoretic cases more likely to be of interest, with both time and space derivatives in the game, there is an added wrinkle to this story: it is often not possible to reduce the order of \emph{all} the derivatives this way, due, for example, to Lorentz contractions. In the point-particle example, reduction of order can effectively be used to remove any terms involving $\ddot x$ or its derivatives,\footnote{Besides, of course, the lowest-order one.} which is to say, any higher-derivative terms. In the case of a scalar EFT like that considered in \cref{sec:u1}, however, this procedure would only suffice to remove terms in the equation of motion involving $\Box\phi$ or its derivatives, which will not include all possible higher-derivative terms. To deal with this one needs simply to do a $3+1$ splitting and reduce the equation of motion to second-order in \emph{time} derivatives, as having higher-order spatial gradients floating around will not induce any Ostrogradski instability; an explicit example of this is shown in \rcite{Eliezer:1989cr}.

\subsubsection{Field redefinitions}

Finally, we consider the question of whether the physical solutions to a higher-derivative EFT can be obtained by solving some other theory with exactly second-order equations of motion. In the case of the point particle the answer is clearly yes: the physical effects of the higher-derivative term in \cref{eq:ppaction-m} could have been entirely accounted for by shifting the mass, $m^2\to(1-\epsilon m^2)m^2$.

This will not be the case in general, but it is important to identify when it is. Consider a scalar-tensor EFT with higher-derivative terms, in which we identify some physical solutions (e.g., cosmological ones) using the reduction-of-order method detailed above. If there turns out to be a theory with second-order equations of motion (i.e., a Horndeski theory) which reproduces those same solutions, then this provides an \textit{a priori} unexpected justification for sticking to the Horndeski class of Lagrangians.
 
The question of whether the physical solutions are also exact solutions to another, healthy theory turns out to reduce to the question of whether unhealthy terms in the action can be removed by \emph{field redefinitions}. It is well-known that effective theories related by (perturbative) field redefinitions describe the same low-energy physics, including in the quantum theory \cite{Georgi:1991ch,Arzt:1993gz}.\footnote{In particular, such perturbative field redefinitions leave the S-matrix unchanged. As discussed in \rcite{Arzt:1993gz}, while the Jacobian induced in the path integral introduces ghost operators, but these ghosts generically either decouple from other fields or are auxiliary.} For example, this was used to greatly simplify the dimension-6 operator basis of the Standard Model EFT \cite{Grzadkowski:2010es} (see also \rcite{Henning:2015alf} for a systematic approach to constructing an EFT operator basis for the Standard Model taking into account field redefinitions).

The freedom to perform perturbative field redefinitions usually means that one can use the lowest-order equations of motion to simplify higher-order operators (although as we will see in \cref{sec:scalar-bottom-up} there are some further subtleties). For instance, taking the point-particle action \eqref{eq:ppaction-m} and applying the field redefinition
\begin{equation}
x\to \tilde x = x + \frac12\epsilon \ddot x \ ,
\end{equation}
we find, after integrating by parts,
\begin{equation}
L = \frac12\left(1+m^2\epsilon\right)\dot{x}^2 - \frac12m^2x^2 + \mo(\epsilon^2) \ ,
\end{equation}
which yields precisely the reduced-order equation of motion \eqref{eq:ppeom-red}. Just as in the reduction-of-order procedure, the field redefinition shunts the higher-derivative terms up to higher orders in the expansion parameter: our original $\epsilon \ddot x^2$ term is now at $\mo(\epsilon^2)$, where we can safely ignore it.

If the utility of the reduction-of-order procedure is to provide a second-order equation of motion which can be solved exactly, then the point of field redefinitions is to identify a separate, manifestly-healthy theory (when it exists) whose \emph{exact} solutions agree, perturbatively, with the physical solutions of the EFT in question.

\subsection{Discussion}

We have presented four equivalent methods for separating out an EFT's physical solutions from its unhealthy, often runaway ones.\footnote{One exception to this equivalence, pointed out in \rcite{Flanagan:1996gw}, arises from the fact that strictly speaking only the equations of motion---not the solutions---are required to be under perturbative control. There can be cases in which corrections to the equations of motion are locally small, but over long times secularly accumulate and lead to large effects on the solution. An example is black hole evaporation, which relies on semiclassical corrections coming to dominate the pure general relativity solution; the semiclassical equations can still be trusted everywhere apart from the singularity and the final stage of evaporation, given that one needs only consider distances much larger than the Planck scale. It is claimed in \rcite{Flanagan:1996gw} that these effects, which are obviously missed by solving perturbatively, are taken into account by the reduction-of-order method. They could potentially also be well-described by open EFTs \cite{Burgess:2014eoa,Burgess:2015ajz}. Whether these sorts of effect are relevant to the cases of interest in this paper, such as modified gravity, we leave to future work.} In practice, either reduction of order or field redefinitions are the most useful techniques in applications.

It is well-understood that when constructing an operator basis for an EFT up to a given order, one writes down all possible terms up to two redundancies: integrations by parts and field redefinitions. It seems reasonable to use these redundancies to organize the operator basis such that it includes as few terms with higher-derivative equations of motion as possible. In the scalar or scalar-tensor context, for instance, this would mean doing one's best to exclude from the basis terms that are not of the generalized galileon or Horndeski type, respectively.

By doing this one identifies what we might call \emph{genuine} higher-derivative terms: terms which lead to higher-order equations of motion and which cannot be removed using any combination of integrations by parts and field redefinitions.\footnote{Or at least cannot be removed without generating further unhealthy terms. Note that higher-order \emph{time} derivatives can always be removed via field redefinitions \cite{GrosseKnetter:1993td}. The question is whether we can remove higher derivatives from the equations of motion while maintaining manifest covariance.} We will find that such genuine higher-derivative terms tend to appear at surprisingly high orders in the EFT expansion. When these genuine higher-derivative operators are present, we advocate for using the reduction of order procedure (possibly after performing a $3+1$ decomposition) to obtain equivalent equations of motion which can then be safely solved.

\section{Higher derivatives in scalar EFTs}
\label{sec:scalar-eft}

In this section, we consider in more detail the question of when higher-derivative terms are and are not genuine, for the example of a real scalar field in both the top-down and bottom-up approaches.

The foils for higher-derivative operators will be the generalized galileons, the most general scalar field terms which lead to second-order equations of motion (on flat space), despite containing second derivatives in the action \cite{Horndeski:1974wa,Deffayet:2011gz}. The generalized galileon Lagrangians in four dimensions are
\begin{align}
\mathcal{L}_2 &= G_2(\phi,X) \ , \nonumber \\
\mathcal{L}_3 &= G_3(\phi,X)\Box\phi \ , \nonumber \\
\mathcal{L}_4 &= G_4(\phi,X)\left[(\Box\phi)^2 - \phi_\mn^2\right] \ , \nonumber \\
\mathcal{L}_5 &= G_5(\phi,X)\left[(\Box\phi)^3 - 3\phi_\mn^2\Box\phi + 2\phi_\mn^3\right] \ , \label{eq:gal-lag}
\end{align}
where $G_n(\phi,X)$ are arbitrary functions of $\phi$ and $X=(\partial\phi)^2$. When $G_n(\phi,X)$ is constant, all of these terms are total derivatives. The standard galileons \cite{Nicolis:2008in} arise when $G_n(\phi,X)\propto X$, in which case $\mathcal{L}_2$ is the canonical kinetic term and $\mathcal{L}_3$, $\mathcal{L}_4$, and $\mathcal{L}_5$ are frequently referred to as the cubic, quartic, and quintic galileons, respectively.

\subsection{The U(1) scalar: genuinely higher derivative?}
\label{sec:u1-redefs}

Let us return to the example of the U(1) scalar discussed in \cref{sec:u1}. The effective action \eqref{eq:cliff-eft-action} up to $\mo(M^{-4})$,
\begin{equation}
\frac{S}{v^2} = \int\dd^4x\left[-\frac12(\partial\theta)^2 + \frac{1}{2M^2}(\partial\theta)^4 - \frac{2}{M^4}\theta_\mn\theta^{\mu\rho}\theta^\nu\theta_\rho + \mo\left(\frac{1}{M^6}\right)\right] \ ,
\end{equation}
contains a healthy term at $\mo(M^{-2})$, as it falls into the generalized galileon class with $G_2(\phi,X)=X^2/2M^2$, but the $\mo(M^{-4})$ term is clearly unhealthy. This problematic higher-derivative term can be written up to a boundary term as
\begin{equation}
\theta_\mn\theta^{\mu\rho}\theta_\rho\theta^\nu \sim \frac12(\partial\theta)^2\left[(\Box\theta)^2 - \theta_\mn\theta^\mn\right] + \theta_\mn \theta^\mu \theta^\nu\Box\theta \ ,
\end{equation}
as can be seen by constructing the total derivative $\partial_\mu\left[(\partial\theta)^2\left(\theta^\mn\theta_\nu-\theta^\mu\Box\theta\right)\right]$. The first two terms make up the \emph{quartic galileon}, which is the generalized galileon with $G_4(\theta,X)\propto X$. The antisymmetric structure of these terms ensures the cancellation of the third- and fourth-order derivatives that appear upon variation \cite{Nicolis:2008in}. The higher-derivative nature is therefore localized to the third and final term, $\theta_\mn\theta^\mu\theta^\nu\Box\theta$. However, this term vanishes on-shell---it is proportional to the lowest-order equation of motion, $\Box\theta$---and can therefore be removed by a field redefinition.

Specifically, by performing the field redefinition
\begin{equation}
\theta \to \tilde\theta = \theta + \frac{2}{M^4}\theta_\mn\theta^\mu\theta^\nu \ ,
\end{equation}
the action \eqref{eq:cliff-eft-action} becomes (after integrations by parts)
\begin{equation}
\frac{S}{v^2} = \int\dd^4x\left[-\frac12(\partial\theta)^2 + \frac{1}{2M^2}(\partial\theta)^4 + \frac{1}{M^4}(\partial\theta)^2\left[\theta_\mn\theta^\mn- (\Box\theta)^2\right] + \mo\left(\frac{1}{M^6}\right)\right] \ .
\end{equation}
The higher-derivative nature of the equations of motion has been pushed off to $\mo(M^{-6})$ and higher, leaving us with galileon dynamics. Indeed, one can check explicitly that solutions to this field-redefined quartic galileon EFT agree, perturbatively, with solutions to both the original EFT and the full theory, $\theta = \theta_0 + \epsilon \theta_1 + \epsilon^2\theta_2$, after performing the field redefinition.

We see in this scalar field example that the EFT physics could in fact have been obtained by considering galileon terms, at least to $\mo(M^{-4})$. As we will see, this behavior is far more prevalent than one might na\"ively expect, providing at least a partial EFT justification for the use of operators with strictly second-order equations of motion. Further, this points the way forward in the search for phenomenologically new higher-derivative EFT operators.

\subsection{Scalar operator basis to dimension 12}
\label{sec:scalar-bottom-up}

We now start to address the question of how often, and under what circumstances, higher-derivative operators in EFTs lead to physical behavior which could not have been considered by restricting from the start to theories with second-order equations of motion, such as the generalized galileons, Horndeski, and Lovelock terms. Specifically, in the context of EFTs for a single scalar field, we will explore what genuinely new terms can be constructed, and whether they are phenomenologically interesting.

We work in a mass-dimension expansion and assume a shift symmetry for simplicity, although dropping the latter assumption would not affect our conclusions. This entails working in the ``bottom-up" approach: our task is to enumerate all of the independent operators at a given mass dimension up to integrations by parts and field redefinitions, determining whether the operator basis is required to have terms that lead to higher derivatives in the equation of motion.

Working up to mass dimension 12, all of the shift-symmetric operators for a single scalar we are allowed to write down, up to boundary terms, are
\begin{align*}
\text{dimension 4:}&& X \\
\text{dimension 5:}&&\text{none} \\
\text{dimension 6:}&& (\Box\phi)^2\\
\text{dimension 7:}&& X\Box\phi\\
\text{dimension 8:}&& X^2,\quad\Box\phi\Box^2\phi\\
\text{dimension 9:}&& \phi_{\mn}^2\Box\phi,\quad (\Box\phi)^3\\
\text{dimension 10:}&& X(\Box\phi)^2,\quad X\phi_\mn^2,\quad\phi_\mn\phi^\mu\phi^\nu\Box\phi, \quad \Box\phi\Box^3\phi\\
\text{dimension 11:}&& \phi_{\mn\alpha}^2\Box\phi,\quad (\partial\Box\phi)^2\Box\phi, \quad (\phi_\mn\partial^\mu\partial^\nu\Box\phi)\Box\phi \\
\text{dimension 12:}&& X^3,\quad (\phi_\mn^2)^2,\quad (\Box\phi)^4,\quad \phi_\mn^3\Box\phi,\quad \phi_\mn^2(\Box\phi)^2,\quad \Box\phi\Box^4\phi
\end{align*}
where we have defined $X\equiv (\partial\phi)^2$. There are many possible ways of writing this list, since we have left out terms which are redundant up to integrations by parts, and have chosen to write things in such a way that as many terms as possible are proportional to $\Box\phi$, which will prove convenient when we apply field redefinitions. The action for the EFT will include all of these terms, suppressed by powers of some scale $\Lambda$ and with arbitrary dimensionless coefficients $c_i$, taken to be order-unity,
\begin{equation}
S = \int\dd^4x\left[-\frac12(\partial\phi)^2 + \frac{c_1}{\Lambda^2}(\Box\phi)^2 + \frac{c_2}{\Lambda^3}(\partial\phi)^2\Box\phi + \frac{1}{\Lambda^4}\left(c_3(\partial\phi)^4 + c_4\Box\phi\Box^2\phi\right) + \cdots\right] \ . \label{eq:scalarbottomupaction}
\end{equation}

Now consider the effect of field redefinitions, working order by order. A field redefinition of the form
\begin{equation}
\phi \to \phi + \frac{1}{\Lambda^n}\psi[\phi,\partial\phi,\partial^2\phi,\cdots]
\end{equation}
will shift the canonical kinetic term by
\begin{equation}
-\frac12(\partial\phi)^2 \to -\frac12(\partial\phi)^2 + \frac{1}{\Lambda^n}\psi\Box\phi - \frac{1}{2\Lambda^{2n}}(\partial\psi)^2 \ ,
\end{equation}
where we have integrated by parts on the second term. Any term which is proportional to $\Box\phi$ in \cref{eq:scalarbottomupaction} at $\mo(\Lambda^{-n})$ can be removed by choosing an appropriate $\psi$ to cancel it out. Because by construction the action contains every term we are allowed to write down, the $(\partial\psi)^2$ term will only shift some of the coefficients of terms already present at $\mo(\Lambda^{-2n})$. We can therefore perform the field redefinition procedure order-by-order, with the effect being that we can simply remove from the operator basis terms proportional to $\Box\phi$.

Performing field redefinitions in this way reduces the operator list significantly, to $\{X,\,X^2,\,X\phi_\mn^2,\,X^3,\,(\phi_\mn^2)^2\}$. The terms $X^n$ only contain first derivatives and so manifestly lead to second-order equations of motion. The other two terms, $X\phi_\mn^2$ and $(\phi_\mn^2)^2$ are higher derivative, although the former is one half of the quartic galileon, $X(\phi_\mn^2-(\Box\phi)^2)$, whose equation of motion is second-order. We can package it into that form with another field redefinition to generate the corresponding $X(\Box\phi)^2$ term, leaving our operator basis in a form which contains no higher-derivative operators until dimension 12:
\begin{align*}
\text{dimension 4:}&& X \\
\text{dimension 5:}&&\text{none} \\
\text{dimension 6:}&& \text{none} \\
\text{dimension 7:}&& \text{none} \\
\text{dimension 8:}&& X^2\\
\text{dimension 9:}&& \text{none} \\
\text{dimension 10:}&& X\left[\phi_\mn^2-(\Box\phi)^2\right] \\
\text{dimension 11:}&& \text{none} \\
\text{dimension 12:}&& X^3,\quad (\phi_\mn^2)^2
\end{align*}

We conclude that the EFT for a shift-symmetric scalar can be put into a fully ghost-free form up to mass dimension 11, with a single higher-derivative operator finally becoming necessary at dimension 12. This is a very high order, far higher than most practical purposes are likely to entail. We see that, while there are no \emph{a priori} obstructions to genuinely higher-derivative terms appearing in EFTs, they may in general not appear until a rather higher order in the expansion than might at first be expected.

Interestingly, in the presence of a mass term, field redefinitions shift $\Box\phi\to m^2\phi$. If we have a dimension-$n$ operator $\Lambda^{n-4}\mo_{n-3}\Box\phi$, then this turns into the dimension-$(n-2)$ operator $\Lambda^{n-4}\mo_{n-3}m^2\phi$. This term appears in the EFT as $\Lambda^{n-7}\phi\mo_{n-3}$ with a coefficient $m^2/\Lambda^2$. If $\Lambda$ is of order the mass of the lightest heavy particle we have integrated out, then by construction $m\ll\Lambda$, and this term that we have generated is similarly suppressed. We see that our conclusions are not qualitatively changed in the absence of a shift symmetry.

\section{Modified gravity as an EFT}
\label{sec:mg-eft}

In this section we consider modified gravity from the effective field theory point of view and consider the role of higher-derivative operators. We will focus on scalar-tensor theories, in which genuinely higher-derivative operators are those which do not fall into the Horndeski class. This class contains four Lagrangians with four free functions $G_n$,
\begin{align}
\mathcal{L}_2 &= G_2(\theta,X) \ , \nonumber \\
\mathcal{L}_3 &= G_3(\theta,X)\Box\theta \ , \nonumber  \\
\mathcal{L}_4 &= G_4(\theta,X)R - 2G_{4,X}\left[(\Box\theta)^2-\theta_\mn^2\right] \ , \nonumber  \\
\mathcal{L}_5 &= G_5(\theta,X)G_\mn\theta^\mn + \frac13G_{5,X}\left[(\Box\theta)^3-3\theta_\mn^2\Box\theta+2\theta_\mn^3\right] \ . \label{eq:horn}
\end{align}
These four Lagrangians are the most general scalar-tensor operators leading to exactly second-order equations of motion. Because they are exactly ghost-free, and because scalar-tensor theories are in a sense minimal modifications of general relativity (as they only add a single, spin-0 degree of freedom), this theory has found heavy use in cosmological applications, e.g., \cite{Amendola:2012ky,Amendola:2016saw}.

It is our aim in this section to investigate where these theories fit in the context of an effective field theory treatment. We will examine this from both the bottom-up and top-down approaches. In the bottom-up case we ask what higher-derivative operators should be considered alongside Horndeski, using as an example a shift-symmetric theory in the derivative expansion. At the four-derivative level there is one term, corresponding to $G_2=X^2$, while at the six-derivative level we find seven terms. Two of these are of the Horndeski form ($G_2=X^3$ and $G_4=X^2$), while the other five (two curvature-only terms and three scalar-tensor terms) are genuinely higher-derivative, and should be included alongside the aforementioned three Horndeski terms when considering EFTs which are organized in a derivative expansion. We remind the reader that for these terms, the ghostly solutions should be removed by using the reduction-of-order technique on the equations of motion.\footnote{When working in the second-order action formalism around, e.g., cosmological backgrounds, one could also use field redefinitions specifically to remove higher time derivatives.} In the top-down case we will consider a specific model of two scalars coupled to gravity---the model of \cref{sec:u1,sec:u1-redefs} with an additional $\Phi^\star\Phi R$ coupling---and integrate out the heavy mode to obtain a ghostly EFT, demonstrating how, in that particular case and to $\mo(M^{-4})$, field redefinitions can be used to package all of the higher derivatives into a Horndeski form.

\subsection{Field redefinitions}

We begin with a technical point, by considering how field redefinitions shift the action. Consider an effective field theory of a scalar field $\phi$ coupled to a metric $g_\mn$, and allow for all possible terms. We will aim to eliminate higher-derivative interactions with field redefinitions of the form
\begin{align}
\phi &\to \phi + \epsilon\psi \ , \nonumber \\
g_\mn &\to g_\mn + \epsilon h_\mn \ , \label{eq:st-redef}
\end{align}
where $\psi$ and $h_\mn$ can depend on the fields and any of their derivatives, and $\epsilon$ is a small parameter which will generically be some inverse power of the cutoff.

The most difficult part of the field redefinition is the terms generated by $h_\mn$, particularly at $\mo(\epsilon^2)$, since we can expect from the start that the $\mo(\epsilon)$ term will be proportional to the Einstein equation.\footnote{In the ``bottom-up" approach the $\mo(\epsilon^2)$ terms will not matter; as in the previous section, we can work order by order and consider only the $\mo(\epsilon)$ part of the transformation. However, when working from the top-down, as we consider below, it is important to ensure that a field redefinition performed at $\mo(\epsilon)$ does not generate new dangerous terms at $\mo(\epsilon^2)$.} We will use the following trick (see, e.g. \cite{Zumalacarregui:2013pma}). Given two metrics $\tilde g_\mn$ and $g_\mn$, the difference between their Christoffel symbols is a tensor, and can therefore be written in a manifestly covariant form,
\begin{align}
\mathcal{K}^\alpha_\mn &\equiv \tilde{\Gamma}^\alpha_\mn - \Gamma^\alpha_\mn\nonumber\\
&= \tilde g^\ab\left(\nabla_{(\mu}\tilde{g}_{\nu)\beta} - \frac12\nabla_\beta \tilde{g}_\mn\right) \ .
\end{align}
Defining the Riemann tensor by $2\tilde\nabla_{[\mu}\tilde\nabla_{\nu]}\equiv\tilde R^\alpha{}_{\beta\mn}v^\beta$, we find that it transforms as
\begin{align}
\tilde R^\alpha{}_{\beta\mn} 
&= R^\alpha{}_{\beta\mn} + 2\tilde\nabla_{[\mu}\mathcal{K}^\alpha_{\nu]\beta} - 2\mathcal{K}^\alpha_{\rho[\mu}\mathcal{K}^\rho_{\nu]\beta} \ .
\end{align}
This allows us to write a simple expression for the transformation of the Einstein-Hilbert term,
\begin{equation}
\sdgt \tilde{R} = \sdgt \tilde{g}^\mn\left(R_\mn - 2\mathcal{K}^\alpha_{\beta[\alpha}\mathcal{K}^\beta_{\mu]\nu}\right) + \sdgt \tilde\nabla_\mu \xi^\mu \ ,
\end{equation}
where $\xi^\mu \equiv \tilde g^\mn \mathcal{K}^\alpha_\mn - \tilde g^{\mu\alpha}\mathcal{K}^\nu_\mn$ packages all the terms with second derivatives of $\tilde g_\mn$ into a total derivative.

Now let us specialize to
\begin{equation}
\tilde g_\mn = g_\mn + \epsilon h_\mn \ ,
\end{equation}
yielding
\begin{align}
\tilde g^\mn &= g^\mn - \epsilon h^\mn + \epsilon^2h^\mu{}_\alpha h^{\alpha\nu} + \mathcal{O}(\epsilon^3) \ , \\
\frac\sdgt\sdg &= 1+ \frac12\epsilon[h] + \frac12\epsilon^2\left(\frac14[h]^2 - \frac12[h^2]\right) + \mathcal{O}(\epsilon^3) \ ,
\end{align}
where we are raising and lowering indices with $g_\mn$ and brackets denote traces. We then have
\begin{equation}
\mathcal{K}^\alpha_\mn = \epsilon g^\ab\left(\nabla_{(\mu}h_{\nu)\beta} - \frac12\nabla_\beta h_\mn\right) + \epsilon^2h^\ab\left(\frac12\nabla_\beta h_\mn-\nabla_{(\mu}h_{\nu)\beta}\right) \ ,
\end{equation}
and a little algebra gives the transformation of the Einstein-Hilbert term, up to a boundary term,
\begin{align}
\frac{\sdgt \tilde{R}}{\sdg} &= R - \epsilon G_\mn h^\mn + \epsilon^2\left[\left(R^\mn-\frac14 Rg^\mn\right)\left(h_{\mu\alpha}h^\alpha{}_\nu-\frac12[h]h_\mn\right) -\frac32\nabla^\mu h^\nu{}_{[\alpha}\nabla_\mu h_{\nu]}{}^\alpha\right] \ . \label{eq:eho2}
\end{align}
Note that in the common case $h_\mn = \Lambda(x)g_\mn$, this becomes simply
\begin{equation}
\frac{\sdgt \tilde{R}}{\sdg} = R + \epsilon\Lambda R + \frac32\epsilon^2(\partial\Lambda)^2 \ . \label{eq:eho2-simp}
\end{equation}

\subsection{Bottom-up: scalar-tensor operator basis}
\label{sec:bottom-up-s-t}

Our goal here is to construct an operator basis for scalar-tensor theories and search for genuinely non-Horndeski terms. We work with a shift-symmetric scalar, although this condition is straightforward to relax. We also assume the absence of any matter coupled to $g_\mn$ besides the scalar field. This is straightforward to include: one simply needs to replace $G_\mn$ by $G_\mn - \Mp^{-2}T_\mn$ in the field redefinition, so that removing higher-derivative terms with field redefinitions will generate contact terms of the form $T_\mn T^\mn$ and so on, as in, e.g., \cite{Bloomfield:2011np}. For certain applications---e.g., to the late Universe---this may be the best way to deal with things. However, all matter can be considered in terms of fields, and so in principle one can also consider field transformations performed on the matter fields. Indeed, this section is precisely concerned with the case in which the matter coupled to gravity is described by a scalar field.

Our EFT action is
\begin{equation}
S = \int\dd^4x\sdg\left[\frac{\Mp^2}{2}R - \frac12(\partial\phi)^2 + \displaystyle\sum_{n=1}^\infty\frac{1}{\Lambda^n}\mo_{n+4}\right] \ ,
\end{equation}
where $\mo_{n+4}$ contains all of the dimension-$(n+4)$ operators we are allowed to write down with arbitrary $\mo(1)$ coefficients. Now let us apply the field redefinitions \eqref{eq:st-redef} with $\epsilon=\Lambda^{-n}$ for some $n$. At $\mo(\epsilon)$ the lowest-order part of the action shifts by
\begin{equation}
\sdg\left(\frac{\Mp^2}{2}R - \frac12(\partial\phi)^2\right) \to \sdg\left[\frac{\Mp^2}{2}R - \frac12(\partial\phi)^2 + \epsilon\left(- \frac{\Mp^2}{2} G_\mn h^\mn +\frac12h^\mn\phi_\mu\phi_\nu -\frac14hX +\psi\Box\phi \right) \right] \ .
\end{equation}
In practice it would be unnecessarily unwieldy to hunt for terms proportional to $G_\mn$ in order to identify the correct $h_\mn$ to remove higher derivatives. We simplify things by defining
\begin{equation}
h_\mn = \bar h_\mn - \frac12 \bar h g_\mn \ , \label{eq:tr-h}
\end{equation}
in which case we have
\begin{equation}
\sdg\left(\frac{\Mp^2}{2}R - \frac12(\partial\phi)^2\right) \to \sdg\left[\frac{\Mp^2}{2}R - \frac12(\partial\phi)^2 + \epsilon\left(- \frac{\Mp^2}{2} R_\mn \bar h^\mn +\frac12\bar h^\mn\phi_\mu\phi_\nu+\psi\Box\phi\right) \right] \ ,
\end{equation}
and so only need to identify terms proportional to $R_\mn$, rather than $G_\mn$. This is because the transformation \eqref{eq:tr-h} from $h_\mn$ to $\bar h_\mn$ is reversible,
\begin{equation}
\bar h_\mn = h_\mn - \frac12 hg_\mn \ ,
\end{equation}
so we can freely specify the $\bar h_\mn$ required to remove a given operator, and then determine the corresponding $h_\mn$ uniquely.

As in the previous section, we use this to identify which higher-derivative terms can be removed using a field redefinition. In this case there are ambiguities in the procedure. Consider a term such as $RX$: do we remove this by using the $R_\mn \bar h^\mn$ term to set $\bar h_\mn\sim Xg_\mn$, or by using the $\bar h^\mn \phi_\mu \phi_\nu$ term with $\bar h_\mn\sim Rg_\mn$? The key point is to look at which additional terms are generated: in the former case we generate $\bar h^\mn \phi_\mu \phi_\nu \sim X^2$, while in the latter we generate $R_\mn \bar h^\mn\sim R^2$. The $R^2$ term should be removed, if possible, as it contains an additional, unphysical degree of freedom,\footnote{While this manages to avoid the Ostrogradski instability, it still introduces new initial conditions.} and doing so would require precisely the same field redefinition which generated this term in the first place, $\bar h_\mn \sim R g_\mn$. Therefore, we should remove a scalar-tensor term like $RX$ using the $R_\mn \bar h^\mn$ part of the transformation. Generalizing this logic suggests the following strategy for dealing with higher-derivative terms in a scalar-tensor EFT:
\begin{enumerate}
\item Remove curvature-only terms using the $R_\mn\bar h^\mn$ coupling.
\item Remove scalar-tensor terms using $R_\mn\bar h^\mn$ again, iterating as necessary.
\item Remove scalar-only terms using $\psi\Box\phi$.
\end{enumerate}

The first two steps generate new terms of the form $\bar h^\mn \phi_\mu \phi_\nu$. While in principle those terms can often be removed using further field redefinitions, this will turn out not to be important in a mass-dimension expansion, as the newly generated terms are both higher-order and suppressed. For example, removing the dimension-6 operator $RX$ generates $X^2$, which is dimension-8. In the bottom-up approach, we write down every operator we are allowed, so these new terms will already be in our operator basis. One might worry, however, that the generated terms come in with a larger coefficient than their counterpart. In fact, the opposite is true, and the generated terms are suppressed. Consider the action at dimension-$(n+4)$,
\begin{equation}
S = \int\dd^4x\sdg\left[\frac{\Mp^2}{2}R - \frac12(\partial\phi)^2 + \cdots + \frac{1}{\Lambda^n}T^\mn R_\mn + \mathcal{O}\left(\frac{1}{\Lambda^{n+1}}\right)\right] \ ,
\end{equation}
where $T_\mn$ is some tensor built out of the fields and their derivatives, and the $\cdots$ includes any terms at lower dimensions which may be present. We can remove the term $T^\mn R_\mn$ by doing a field redefinition with
\begin{equation}
\bar h_\mn = \frac{2}{\Mp^2} T_\mn \ ,
\end{equation}
corresponding to
\begin{equation}
g_\mn \to g_\mn + \frac{2}{\Lambda^n\Mp^2}\left(T_\mn-\frac12Tg_\mn\right) \ .
\end{equation}
The action then becomes
\begin{equation}
S = \int\dd^4x\sdg\left[\frac{\Mp^2}{2}R - \frac12(\partial\phi)^2 + \frac{1}{\Mp^2\Lambda^n}T^\mn \phi_\mu\phi_\nu + \mathcal{O}\left(\frac{1}{\Lambda^{n+1}}\right)\right] \ .
\end{equation}
The new term, $T^\mn \phi_\mu \phi_\nu$, is dimension-$(n+6)$, so according to the bottom-up EFT logic another term of this form will appear in the action suppressed by $\Lambda^{-(n+2)}$ with a generically $\mo(1)$ coefficient $c_{n+2}$,
\begin{align}
\frac{1}{\Mp^2\Lambda^n}T^\mn\phi_\mu\phi_\nu + \frac{c_{n+2}}{\Lambda^{n+2}}T^\mn\phi_\mu\phi_\nu &= \frac{1}{\Lambda^{n+2}}T^\mn\phi_\mu\phi_\nu\left(\frac{\Lambda^2}{\Mp^2}+c_{n+2}\right) \nonumber \\
&\approx \frac{c_{n+2}}{\Lambda^{n+2}}T^\mn\phi_\mu\phi_\nu \ .
\end{align}
The term that we generated with our field redefinition is suppressed compared to its already-present counterpart as long as the EFT cutoff is sub-Planckian and the coefficient $c_{n+2}$ is order unity. So we can ignore $\bar h_\mn\phi^\mu\phi^\nu$ terms generated in steps 1 and 2 of the above procedure: they are much smaller than terms that we would write down two orders higher in the mass-dimension expansion.

We see that any term proportional to $R_\mn$ and $\Box\phi$ (again, assuming shift symmetry) can be removed by a field redefinition. This implies that the only genuinely higher-derivative scalar-tensor operators---i.e., those which are not in the Horndeski class and which contribute new dynamics---are those which involve the Riemann tensor.

With this in mind, let us enumerate an example of a scalar-tensor operator basis. We work in the derivative expansion\footnote{This is in contrast to the approach in \cref{sec:scalar-bottom-up}, where we expanded in terms of mass dimension. Different expansions are useful for different purposes---for example, the derivative expansion is used in the EFT of inflation \cite{Weinberg:2008hq}, and also emerged naturally in the complex scalar example considered in \cref{sec:u1}, as that was the EFT for the Goldstone boson of a spontaneously-broken U(1) symmetry. Moreover, the operator list at the six-derivative level is somewhat more compact than its mass-dimension counterpart, and allows for more focus on non-Horndeski scalar-tensor couplings than on pure curvature terms. For example, in the six-derivative operator basis we have mass dimension-8 and dimension-10 scalar-tensor terms, and a dimension-12 scalar-only term. If we were to include all possible terms of the same order, we would have a veritable zoo of pure curvature couplings; for instance, we would find terms of the form $\mathrm{Riem}^4$ and $\mathrm{Riem}^2\nabla\mathrm{Riem}$ at dimension 8 and $\mathrm{Riem}^5$ and a host of others at dimension 10.} and assume shift-symmetry.\footnote{The condition of shift symmetry is straightforward to relax, in which case one ends up with a six-derivative extension of the EFT-of-inflation operator basis presented in \rcite{Weinberg:2008hq}. The application of such a basis to inflation is work in progress \cite{inprep}.} The operators involving only the scalar field are precisely those listed up to mass dimension 12 in \cref{sec:scalar-bottom-up}: all of the terms in that list have four or six derivatives, and conversely there cannot be shift-symmetric six-derivative scalar operators which have mass dimension greater than 12. While in principle one can write down new scalar-only terms compared to that list (due to the fact that we are now on curved space, so covariant derivatives do not commute), we can always write such new terms in terms of those already considered in the flat-space list plus curvature couplings. Moreover, any operator involving the Ricci scalar or tensor can be removed with a metric field redefinition. Our task is therefore to determine all six-derivative operators involving the Riemann tensor which are not related by total derivatives or field redefinitions.

Couplings which necessarily\footnote{By ``necessarily" we mean even after taking into account all possible integrations by parts.} involve the Riemann tensor are both novel, as they do not fall into the Horndeski class; and tricky to enumerate properly, as the Riemann tensor comes with a variety of redundancies. One needs to take into account the symmetries of the Riemann tensor and the Bianchi identities, as well as the so-called dimensionally-dependent identities \cite{Edgar:2001vv}. The xTras \cite{Nutma:2013zea} tensor algebra package for \textit{Mathematica} has a number of tools which help to automate the task of accounting for these redundancies. We begin by enumerating all possible tensorial structures involving covariant derivatives of $\phi$, at least one copy of the Riemann tensor (since scalar-only terms were completely enumerated in \cref{sec:scalar-bottom-up}), and covariant derivatives of Riemann. We then construct all of the index contractions for each structure, up to Bianchi identities and dimensionally-dependent identities. Finally we look for redundancies up to boundary terms. At every step, we freely throw away any term involving $R_\mn$, $R$, or $\Box\phi$, as such terms can be removed using field redefinitions. We are left with the following operator basis:
\begin{align}
\text{Four derivatives:}&& X^2 \nonumber \\
\text{Six derivatives:}&& X^3, \quad X\left[\phi_\mn^2-(\Box\phi)^2 + \frac14XR\right], \nonumber \\
&&R_{\mn\ab}\phi^{\mu\alpha}\phi^{\nu\beta}, \quad R_{\mn\ab}\phi^\mu\phi^\alpha\phi^{\nu\beta}, \quad  XR_{\mn\ab}^2,  \nonumber \\
&&R^\mn{}_\ab R^\ab{}_{\rho\sigma}R^{\rho\sigma}{}_\mn, \quad (\nabla R_{\mn\ab})^2
\end{align}
We have organized the six-derivative list into three lines: terms in the Horndeski class which lead to second-order equations of motion, non-Horndeski scalar-tensor couplings, and curvature-only operators, respectively. Note that as in \cref{sec:scalar-bottom-up}, we have used a further field redefinition to massage the operator $X\phi_\mn^2$ into quartic Horndeski form with $G_4\sim X^2$.

We emphasize that in addition to the two six-derivative Horndeski operators, we have identified five new, genuinely higher-derivative operators which should be considered alongside their Horndeski counterparts in phenomenological applications.

\subsection{Top-down: emergent Horndeski from a U(1) scalar}

We now turn to higher derivatives in the top-down approach. We study an example ``UV" theory\footnote{We put ``UV" in quotes because this theory is non-renormalizable; it is more a UV \emph{extension} than a UV completion of our scalar-tensor EFT.} whose low-energy EFT includes a number of higher-derivative terms at both the four- and six-derivative level, and then show how field redefinitions can consistently be used to remove these, leaving us solely with Horndeski terms. This serves to illustrate how EFTs that appear to have higher-derivative terms can nevertheless often be described by the special classes of theories with exactly second-order equations of motion.

\subsubsection{Integrating out the heavy field}

Consider the complex scalar model introduced in \cref{sec:u1} and couple it to gravity in the minimal way,\footnote{This theory is not renormalizable and so should itself be seen as an effective theory. In this respect, the $\Phi^\star\Phi R$ coupling is the lowest-order one which respects the symmetries, diffeomorphism and U(1) invariance.} through a $\Phi^\star\Phi R$ coupling,
\begin{equation}
S = \int\dd^4x \sdg\left(\frac{\Mp^2}{2}R + \xi\Phi^\star\Phi R - \partial_\mu\Phi^\star\partial^\mu\Phi - V(\Phi^\star\Phi) + \mathcal{L}_\mathrm{m}(g)\right) \ , \label{eq:UV-ST}
\end{equation}
with the same $V(\Phi^\star\Phi)$ as above. Splitting $\Phi$ into real and imaginary pieces as before, $\Phi = \frac{v}{\sqrt{2}}(1+\rho)e^{i\theta}$, the action can be written
\begin{equation}
\frac{S}{v^2} = \int\dd^4x\sdg\left[\frac12\left(\frac{\Mp^2}{v^2}+\xi(1+\rho)^2\right)R-\frac12(\partial\rho)^2 - \frac12(1+\rho)^2(\partial\theta)^2 - V(\rho) + \frac{1}{v^2}\mathcal{L}_\mathrm{m}(g)\right] \ .
\end{equation}

We integrate $\rho$ out following \cite{Burgess:2014lwa}. Varying the action with respect to $\rho$ we find
\begin{equation}
\Box\rho + \xi(1+\rho)R - (1+\rho)(\partial\theta)^2 - V'(\rho) = 0 \ .
\end{equation}
We can obtain a (highly non-local and recursive) solution for $\rho$ by introducing an appropriate Green's function,
\begin{equation}
\rho(x') = -\int\dd^4x G(x,x')\left\{\left[1+\rho(x)\right]\left[\partial\theta(x)\right]^2 + V_\mathrm{int}'\left[\rho(x)\right] - \xi\left[1+\rho(x)\right]R(x)\right\} \ , \label{eq:s-t-rho-eom}
\end{equation}
where $V_\mathrm{int}\equiv V - \frac12M^2\rho^2$ and the Green's function is given by
\begin{equation}
(-\Box+M^2)G(x,x') = \delta^{(4)}(x-x') \ . \label{eq:green}
\end{equation}
The EFT for $\theta$ and $g_\mn$ is obtained by substituting this expression for $\rho$ into the action. Of course, the resultant action would be entirely useless. The power of effective field theory comes from utilizing the largeness of $M$. This renders our solution for $\rho$ local and, in the process, introduces the apparently-ghostly interactions. We expand $\rho$ and $G$ in powers of $1/M$ as
\begin{equation}
\rho(x) = \displaystyle\sum_{n=1}^\infty\frac{r_n(x)}{M^{2n}},\qquad G(x,x') = \displaystyle\sum_{n=1}^\infty\frac{g_n(x,x')}{M^{2n}} \ .
\end{equation}
Plugging these definitions into \cref{eq:green} we find
\begin{equation}
g_n = \Box^{n-1}\delta^{(4)}(x-x') \ ,
\end{equation}
and then using these in \cref{eq:s-t-rho-eom} we obtain, working to $\mo(M^{-4})$ \ ,
\begin{align}
r_1&=\xi R-(\partial\theta)^2 \ , \\
r_2 &= -\frac12\left[\xi R-(\partial\theta)^2\right]^2 + \Box\left[\xi R-(\partial\theta)^2\right] \ .
\end{align}
The action, up to $\mathcal{O}(M^{-4})$, is then
\begin{align}
\frac{S}{v^2} &\simeq \int\dd^4x\sdg\bigg[\frac{\Meff^2}{2v^2}R - \frac12(\partial\theta)^2 + \frac{1}{M^2}\left(\frac{\xi^2}{2}R^2 - \xi XR + \frac12(\partial\theta)^4\right) \nonumber \\
&\hphantom{{}\simeq} +\frac{1}{M^4}\left(\frac{\xi^2}{2}R\Box R - \xi R\Box X -2\theta_\mn\theta^{\mu\rho}\theta_\rho\theta^\nu\right) + \frac{1}{v^2}\mathcal{L}_\mathrm{m}\bigg] \ , \label{eq:s-t-eft-action}
\end{align}
where the effective Planck mass is $\Meff^2 \equiv \Mp^2 + \xi v^2$. For convenience, we now write the coefficient of the Einstein-Hilbert term as $\Lg \equiv \frac{\Meff^2}{2v^2}$. Note that for our purposes $\theta_\mn$ can be defined with either partial or covariant derivatives, since it appears in the combination $\theta_\mn\theta^{\mu\rho}\theta_\rho\theta^\nu=\frac14(\partial X)^2$.

\subsubsection{Eliminating higher derivatives}

The action \eqref{eq:s-t-eft-action} contains a variety of terms which will lead to higher-order equations of motion: indeed, this is the case for every single term at $\mo(M^{-2})$ and $\mo(M^{-4})$ besides $(\partial\theta)^4$, as none of them fall into the Horndeski class \eqref{eq:horn}.\footnote{Of course, $R^2$ is secretly of the Horndeski form, as it can be written in terms of a (healthy) scalar-tensor theory after a conformal transformation. However, that scalar has no relation to $\theta$ or $\rho$, and therefore should not appear in physical solutions; while it is not ghostly, it is nevertheless a spurious degree of freedom which will lead to extra solutions that are non-perturbative in $1/M$ \cite{Simon:1991bm}.} Our aim is to use field redefinitions to remove as many of the non-Horndeski terms as possible; we will find that, in fact, all of them are redundant, leaving us with just $\mathcal{L}_2$ and $\mathcal{L}_4$ Horndeski terms.

To do the field redefinitions we follow the strategy outlined in the bottom-up case: start by using metric redefinitions to remove curvature-only terms like $R^2$, then use metric redefinitions again to eliminate scalar-tensor couplings through the $\bar h_\mn \phi^\mu \phi^\nu$ transformation of the action, and finally use a scalar redefinition to eliminate problematic scalar-only operators. We emphasize that this strategy really is best suited for the bottom-up approach. A given term can often be removed using different field redefinitions, which generate different operators at higher orders. For instance, an $\mo(M^{-2})$ field redefinition will generate new terms at $\mo(M^{-4})$, and these terms will in general be different depending on whether we remove a scalar-tensor coupling through $\bar h_\mn R^\mn$ or $\bar h_\mn \phi^\mu \phi^\nu$. In the bottom-up approach, every allowed term is already present, and so we are not particularly concerned about which such terms are generated. In the top-down case, some of the allowed operators are absent, so the wrong choice for the field redefinitions might introduce unwanted new terms that are difficult or impossible to exorcise. The safest approach is to write down the most general possible field redefinition and then pick its coefficients in order to remove all of the unwanted terms. In the relatively simple model we look at in this section, however, this turns out to be unnecessary, and picking the simplest field redefinitions will be sufficient to remove all non-Horndeski terms to $\mo(M^{-4})$.

Let us start with the $\mo(M^{-2})$ terms in \cref{eq:s-t-eft-action}. Recall from \cref{sec:bottom-up-s-t} that field redefinitions of the scalar and metric shift the action by terms proportional to $R_\mn$ and $\Box\theta$, respectively. The $R^2$ and $XR$ terms are both proportional to $R_\mn$. A field redefinition
\begin{equation}
g_\mn\to g_\mn + \frac{1}{M^2}h_\mn
\end{equation}
will shift these terms by
\begin{equation}
\frac{\xi^2}{2}R^2 - \xi XR\to \frac{\xi^2}{2}R^2 - \xi XR - \Lg R_\mn\bar h^\mn + \frac12 \bar h^\mn \theta_\mu \theta_\nu \ ,
\end{equation}
with $\bar h_\mn = h_\mn - \frac12hg_\mn$. Starting with curvature-only pieces, we can eliminate the unwanted $R^2$ term by choosing $\bar h_\mn \propto R g_\mn$. This generates an additional term $\bar h^\mn \theta_\mu \theta_\nu \propto XR$, renormalizing the coefficient of the $XR$ term. We can then remove this term by adding to the field redefinition a piece $\bar h_\mn \propto X g_\mn$. Working through these steps explicitly, the final field redefinition is
\begin{equation}
h_\mn = -\bar h_\mn = \left[\bar\xi\left(1-\frac{\bar\xi}{4}\right)X-\frac{\bar\xi^2}{2}\lambda_g R\right] g_\mn \ , \label{eq:m2redef}
\end{equation}
where we have defined
\begin{equation}
\bar\xi \equiv \frac{\xi}{\lambda_g} = 2\left(1+\frac{\Mp^2}{\xi v^2}\right)^{-1} \ .
\end{equation}
Note that if $\xi v^2\ll\Mp^2$ then $\bar\xi\ll1$. This leaves us with the action
\begin{equation}
\frac{S}{v^2} = \int\dd^4x\sdg\left[\lambda_g R - \frac12(\partial\theta)^2 + \frac{(2-\bar\xi)^2}{8M^2}X^2 + \frac{1}{M^4}\mathcal{L}_{M^{-4}} + \mo\left(\frac{1}{M^6}\right)\right] \ ,
\end{equation}
where the $\mathcal{O}(M^{-4})$ contribution is, up to boundary terms,
\begin{equation}
\mathcal{L}_{M^{-4}} = \left(\frac{\xi^2}{2} + \frac98\frac{\xi^4}{\Lg}\right)R\Box R - \xi\left(1 + 3\frac{\xi^2}{\Lg} - \frac38\frac{\xi^3}{\Lg^2}\right)R\Box X - \left(2 + 6\frac{\xi^2}{\Lg} - \frac38\frac{\xi^4}{\Lg^3}\right)\theta_\mn\theta^{\mu\alpha}\theta^\nu\theta_\alpha \ . \label{eq:action-redef-m2}
\end{equation}
Note that the $\mo(M^{-4})$ piece receives contributions from the $\mo(M^{-2})$ field redefinition \eqref{eq:m2redef}. These are of two types. There are higher-order corrections from the $\mo(M^0)$ terms, i.e., the Einstein-Hilbert and canonical kinetic pieces, using \cref{eq:eho2} or \cref{eq:eho2-simp} to calculate the former. Then there are $\mo(M^{-4})$ corrections to the original $\mo(M^{-2})$ terms in the action \eqref{eq:s-t-eft-action}. As it turns out, all of these corrections simply shift the coefficients of $\mo(M^{-4})$ terms in the action \eqref{eq:s-t-eft-action} after some integrations by parts.

We highlight again that the field redefinition we chose was not unique; indeed, any field redefinitions of the form
\begin{align}
\theta &\to \theta + \frac{1}{M^2}\left(a_1X -b_4\lambda_gR\right) \ , \\
g_\mn &\to g_\mn + \frac{1}{M^2}\left\{\left[-\frac{\bar\xi^2}{2}\lambda_g R + \bar\xi\left(1-\frac{\bar\xi}{4}\right)X + b_4\Box\theta\right] g_\mn + b_5\theta_\mu\theta_\nu + b_6\theta_\mn\right\} \ ,
\end{align}
will remove higher-derivative interactions at $\mo(M^{-2})$, leaving us with
\begin{eqnarray}
\frac{S}{v^2} = \int\dd^4x\sdg\left\{\lambda_g R \right. &-& \frac12(\partial\theta)^2 + \frac{1}{2M^2}\left[\left(1+\frac{b_5}{2}-\bar\xi+\frac{\bar\xi^2}{4}\right)X^2 + \left(2a_1-b_4-b_6\right)X\Box\theta \right.\nonumber \\
&-& \left.\left.  2\bar b_5G_\mn\theta^\mu\theta^\nu\right] + \mo\left(\frac{1}{M^{4}}\right)\right\} \ .
\end{eqnarray}
In addition to the $X^2$ term found above, this includes the cubic Horndeski with $G_3\propto X$ and the quartic Horndeski with $G_4\propto X$, which (up to boundary terms) can be written as $G_\mn \theta^\mu\theta^\nu$. In principle one should leave the parameters $a_1$, $b_{4,5,6}$ free in case particular choices are necessary to avoid inducing higher-derivative terms at higher orders in the EFT. In our case, we will see that the simplest choice $a_1=b_4=b_5=b_6=0$ suffices. 

Next we use another set of field redefinitions to remove the unhealthy terms at $\mo(M^{-4})$ in \cref{eq:action-redef-m2}. Each of the three terms in $\mathcal{L}_{M^{-4}}$ is not of the Horndeski form, although, much like this model's flat-space version discussed in \cref{sec:u1-redefs}, the scalar interaction $\theta_\mn\theta^{\mu\alpha}\theta^\nu\theta_\alpha$ secretly contains a Horndeski term,
\begin{equation}
\theta_\mn\theta^{\mu\alpha}\theta^\nu\theta_\alpha \sim -\frac12\mathcal{L}_{G4} + \theta_\mn\theta^\mu\theta^\nu\Box\theta - \frac12X \left(R_\mn - \frac14 R g_\mn\right) \theta^\mu\theta^\nu \ , \label{eq:secret-horndeski}
\end{equation}
where
\begin{equation}
\mathcal{L}_{G4} = \frac14X^2R + X\left[\theta_\mn^2 - (\Box\theta)^2\right] \ .
\end{equation}
is the quartic Horndeski Lagrangian with $G_4 = X^2/4$. This is due to the total derivative combination
\begin{align}
\nabla_\mu\left[X\left(\theta^\mn\theta_\nu - \theta^\mu\Box\theta\right)\right] &= 2\theta_\mn\theta^{\mu\alpha}\theta^\nu\theta_\alpha - 2\theta_\mn\theta^\mu\theta^\nu\Box\theta + X R_\mn \theta^\mu\theta^\nu + X\left[\theta_\mn^2 - (\Box\theta)^2\right] \\
&= \mathcal{L}_{G4} + 2\theta_\mn\theta^{\mu\alpha}\theta^\nu\theta_\alpha - 2\theta_\mn\theta^\mu\theta^\nu\Box\theta + X \left(R_\mn - \frac14 R g_\mn\right) \theta^\mu\theta^\nu \ ,
\end{align}
Once again we consider field redefinitions of the form
\begin{align}
g_\mn &\to g_\mn + \frac{1}{M^4}h_\mn \ , \\
\theta &\to \theta + \frac{1}{M^4}\psi \ .
\end{align}
The (simplest) field redefinitions which serve to remove the ghostly terms are straightforward to work out, because they do not end up generating new terms that are not already present. Removing $R\Box R$ with $\bar h_\mn \sim \Box Rg_\mn$ generates $\bar h^\mn \theta_\mu\theta_\nu \sim X\Box R$, which simply renormalizes one of our coefficients. We then remove $X\Box R$ with $\bar h_\mn \sim \Box X g_\mn$, generating $\bar h^\mn \theta_\mu \theta_\nu \sim X \Box X$, which up to a boundary term is equivalent to $(\partial X)^2 = 4\theta_\mn \theta^{\mu\alpha}\theta_\alpha \theta^\nu$, which again just renormalizes one of the coefficients in \cref{eq:action-redef-m2}. This leaves us with a term of the form \eqref{eq:secret-horndeski}. The term proportional to $\Box\theta$ can be removed with $\psi\sim\theta_\mn\theta^\mu\theta^\nu$. The final term can be dealt with in two ways: by setting $\bar h_\mn \propto X(R_\mn - \frac14 R g_\mn)$ and by using the $\bar h_\mn \theta^\mu\theta^\nu$ part of the transformation, or by setting $\bar h_\mn \propto X(\theta_\mu\theta_\nu - \frac14X g_\mn)$ and by using $\bar h_\mn R^\mn$.\footnote{Notice that $(R_\mn - \frac14 Rg_\mn)\theta^\mu\theta^\nu = (\theta_\mu\theta_\nu-\frac14Xg_\mn)R^\mn$.} The former would introduce unwanted curvature-squared terms of the type we have just removed, while the latter generates a perfectly benign $X^3$ term.

We therefore choose a field redefinition such that $\bar h_\mn$ includes terms proportional to $\Box R g_\mn$, $\Box X g_\mn$, and $X(\theta_\mu\theta_\nu-\frac14Xg_\mn)$, as well as $\psi\propto\theta_\mn\theta^\mu\theta^\nu$ to vacuum up the remaining $\theta_\mn\theta^\mu\theta^\nu\Box\theta$ term in \cref{eq:secret-horndeski}. After some algebra to determine their coefficients, and transforming from $\bar h_\mn$ back to $h_\mn$, the appropriate field redefinition is
\begin{align}
h_\mn &= \left[-\frac12\Lg\bar\xi^2\left(1 + \frac94\Lg\bar\xi^2\right)\Box R + \bar\xi\left(1 - \frac{\bar\xi}{4} + 3\Lg\bar\xi^2 - \frac{15}{16}\Lg\bar\xi^3\right)\Box X\right]g_\mn \nonumber\\
&\hphantom{{}=} + \frac{(2-\bar\xi)^2(1+3\Lg\bar\xi^2)}{4\Lg}X\left(\theta_\mu\theta_\nu - \frac14Xg_\mn\right) \ , \\
\psi &= \frac{(2-\bar\xi)^2(1+3\Lg\bar\xi^2)}{2}\theta_\mn\theta^\mu\theta^\nu \ .
\end{align}
Although we began at a rather nontrivial starting point, this procedure leaves us with the perfectly healthy, and remarkably simple, final action,
\begin{equation}
\frac{S}{v^2} = \int\dd^4x\sdg\left[\lambda_g R - \frac12(\partial\theta)^2 + \frac{(2-\bar\xi)^2}{8M^2}(\partial\theta)^4 + \frac{(2-\bar\xi)^2(1+3\Lg\bar\xi^2)}{4M^4}\left(\mathcal{L}_{G4} + \frac{3}{8\Lg}(\partial\theta)^6\right) + \mo\left(\frac{1}{M^6}\right)\right] \ .
\end{equation}
Note that, in terms of the original parameters in the action, the coefficients above are
\begin{align}
2 - \bar\xi &= \frac{2\Mp^2}{\Mp^2+\xi v^2} \ , \\
1+3\Lg\bar\xi^2 &= 1 + 6\xi\frac{\xi v^2}{\Mp^2+\xi v^2} \ .
\end{align}

We conclude this section with a few points of interest about this example. First, we note that in the flat-space limit ($\bar\xi\to0$, $\Lg\to\infty$) we obtain the same result as earlier. The $X^3$ term is new and results from including gravity. Note also that the strong-coupling scale is not necessarily $M$, and in fact is different for different terms. To see this, we canonically normalize, $\theta_\mathrm{c}=v\theta$, finding different scales suppressing each of the remaining operators: $\sqrt{Mv}=\lambda^{1/4}v$ for the $(\partial\theta_\mathrm{c})^2$ term, $(M^2v)^{1/3}=\lambda^{1/3}v$ for $\mathcal{L}_{G4}(\theta_\mathrm{c})$, and $(M^2v\Mp)^{1/4} = \lambda^{1/4}(v^3\Mp)^{1/4}$ for $(\partial\theta_\mathrm{c})^6$. In the second equalities we have used $M^2=\lambda v^2$, where unitary should require $\lambda\lesssim1$. Note that only the $X^3$ term has a cutoff that depends on $\Mp$, consistent with the flat-space limit.

Finally, let us remark briefly on the seemingly-contradictory messages of this section and the one preceding it. In the bottom-up discussion, we showed that a scalar-tensor EFT generically contains higher-derivative interactions which should be included alongside Horndeski terms of the same size (e.g., mass dimension or number of derivatives). Here we have shown in an explicit example an EFT of the same type considered in the bottom-up approach---up to six-derivative level in a derivative expansion with shift symmetry---and found that the non-Horndeski terms do not appear.

In a certain sense this result is not at all surprising, given our starting point \eqref{eq:UV-ST}. The only scalar-tensor coupling we considered in the UV coupling was through the Ricci scalar, $\Phi^\star\Phi R$, so there is no way to generate the Riemann tensor couplings which we argued are the only genuinely higher-derivative ones.\footnote{In principle Riemann couplings could have arisen from commuting covariant derivatives in the scalar sector; for example, the six-derivative term $R_{\mn\ab}\phi^{\mu\alpha}\phi^\nu\phi^\beta$, which we have shown is genuinely higher derivative, is equivalent to the scalar-only term $2\phi_{[\mn]\alpha}\phi^{\mu\alpha}\phi^\nu$.} The question of whether our ``UV" theory should have Riemann couplings to $\Phi^\star\Phi$ is best addressed by viewing that theory as itself an EFT. Riemann couplings respecting the U(1) symmetry will be irrelevant and not appear until a higher order in the EFT, thus protecting the lower-energy $\theta$ EFT from genuine higher derivatives. This is a mechanism by which UV physics might conspire to leave the low-energy EFT with purely second-order terms.

\section{Conclusions}
\label{sec:conc}

We have performed a detailed study of the role of higher-derivative operators in effective field theories, focusing on scalar fields with and without gravity. In \cref{sec:higher-derivs} we reviewed various methods of extracting physical solutions to EFTs with higher-order equations of motion, and in \cref{sec:scalar-eft,sec:mg-eft} applied these to scalar fields in both top-down and bottom-up approaches. In particular, we showed that higher-derivative operators may often be left out of the operator basis until fairly high orders in a mass-dimension expansion due to the ability to make perturbative field redefinitions; these can frequently be used to remove higher-derivative operators or package them into special forms like the galileon or Horndeski classes, operators which have second-order equations of motion despite also having second derivatives in the action.

We also identified in these cases the most relevant genuine (i.e., not removable via field redefinition) higher-derivative operators that should be included. These terms may not be physically special---the associated Ostrogradski ghost is an artifact of the EFT truncation and does not lead to a physical instability, as is well-known---but present a challenge when solving equations numerically, as the ghost which lives near the cutoff nevertheless can strongly impact solutions to the equations of motion. When it is possible to use field redefinitions to package such terms into special second-order forms, this difficulty can be avoided entirely and the resultant equations of motion can be solved. When it is not, we have explicitly described how to perturbatively reduce the order of the equations of motion, as discussed in \cref{sec:reduction}.

The EFT approach we have described in this paper is well-understood in typical particle physics models, and its applicability is clearly defined. Applied to some of the cosmological models we have discussed here, it allows us to rigorously understand small corrections to general relativistic dynamics, which could be probed, for instance, in precision cosmological observations. During the inflationary r\'egime, this can lead to potentially-testable new effects \cite{inprep}. If the EFT is relevant in the late Universe, then these corrections are a worthwhile target for next-generation probes. In this sense, we would be testing modified gravity in a way analogous to physics beyond the Standard Model: using precision measurements to determine the scale of new gravitational physics and bound (or even measure) the various EFT coefficients.

This is an important use of the method, but it is worth pointing out that it is nevertheless less ambitious than the uses to which modified gravity theories are more often put, such as driving inflation or late-time acceleration. That approach relies on nonrenormalizable terms dominating over their lower-order counterparts, which, in some cases, can place the validity of the EFT in serious jeopardy. A particularly notable example of this is Starobinsky inflation \cite{Starobinsky:1980te}, which generates an inflationary epoch by adding an $R^2$ term to the Einstein-Hilbert action. This term is one of the first to arise in an EFT expansion of gravity, as discussed in \cref{sec:qg}, but the inflationary solutions, which require the $R^2$ term to dominate over Einstein-Hilbert, are not physical solutions to the EFT in the sense discussed in this paper \cite{Simon:1991bm}, and indeed the curvature-cubed and higher terms in the EFT expansion which would become important alongside the $R^2$ term spoil inflation \cite{Burgess:2016owb}.

Attempts to use actions such as $f(R)$ or Horndeski to drive an accelerating phase face a similar challenge in justifying their validity from the standpoint of effective field theory. in such applications, it is important to explore  whether there are reasons for irrelevant operators to dominate over canonical ones without leaving the EFT, so that the particular term or set of terms used may remain large while all of the other (infinite) terms in the EFT expansion may be safely chosen to be small.

These are not, of course, insurmountable obstacles, but rather provide an opportunity to sharpen our thinking about the modified gravity theories we test against experiment. Fortunately there already exist examples to guide us, in which EFT effects can lead to large changes in the solutions. One possibility is to reorganize the EFT expansion so that the relative sizes of terms are determined not by their mass dimensions but, e.g., the number of derivatives per field.\footnote{See however \rcite{Kaloper:2014vqa} for counterarguments.} An example of this class is the DBI action, whose Lagrangian is of the form $\mathcal{L}_\mathrm{DBI}\sim-\Lambda^4\sqrt{1+X/\Lambda^4}$ \cite{deRham:2010eu}. In the ultra-relativistic limit $|X|\sim1$, every term in the expansion is important, but the equation of motion ensures that $\partial^2\phi$ remains small, allowing the EFT to remain under control. Moreover, the DBI action possesses an additional symmetry, $\phi\to\phi+v_\mu x^\mu + \phi v^\mu \phi_\mu/\Lambda^4$, which sets the coefficients of each of the infinite number of terms in the expansion in $X$. This type of EFT reorganization also applies to the galileons, which contain their own additional (galilean) symmetry and possess a non-renormalization theorem which prevents large higher-derivative corrections from being generated \cite{Luty:2003vm,Hinterbichler:2010xn,Goon:2016ihr}. This has been claimed to hold for more general $P(X)$ theories \cite{deRham:2014wfa}. In the inflationary context this applies to, for example, ghost inflation \cite{ArkaniHamed:2003uy,ArkaniHamed:2003uz}, where the leading correction is the term with only a single two-derivative piece, i.e., an $X$-dependent cubic galileon, $Q(X)\Box\phi$, which is suppressed compared to $P(X)$ by $H/\Lambda$ \cite{Weinberg:2008hq}. In such cases the role of higher-derivative operators, dealt with as described in this paper, bears investigation.

\begin{acknowledgments}
We thank Ana Ach\'ucarro, Yashar Akrami, Sina Bahrami, Daniel Baumann, Lasha Berezhiani, Cliff Burgess, Claudia de Rham, Sergei Dubovsky, \'{E}anna Flanagan, Gregory Gabadadze, Ruth Gregory, Brian Henning, Austin Joyce, Scott Melville, Tony Padilla, Jeremy Sakstein, Alessandra Silvestri, David Stefanyszyn, and Andrew Tolley for useful discussions. We are grateful to Cliff Burgess and Scott Melville for comments on a draft. A.R.S. is supported by funds provided by the Center for Particle Cosmology.  The work of M.T was supported in part by US Department of Energy (HEP) Award DE-SC0013528, and by NASA ATP grant NNX11AI95G. 
\end{acknowledgments}

\bibliography{bibliography}

\end{document}